\documentclass[
reprint,
nofootinbib,
 amsmath,amssymb,
 aps,
]{revtex4-2}
\usepackage{graphicx}  
\usepackage{dcolumn}   
\usepackage{bm}        
\usepackage{amssymb}   
\usepackage{xcolor}
\usepackage{amsmath}
\usepackage[normalem]{ulem}
\usepackage{graphicx}
\usepackage[colorlinks=true,linktocpage=true,linkcolor=blue,citecolor=blue]{hyperref}
\newcommand{\Chi}{\chi}
\preprint{}

\begin{document}

\title{Elliptic Anisotropy from Quantum Diffraction}

\author{Erik Carri\'o}
\email{erikcarrioubeda@gmail.com}
\author{Daniel Pablos}
\email{pablosdaniel@uniovi.es}
\affiliation{Departamento de F\'isica, Universidad de Oviedo, Avda. Federico Garc\'ia Lorca 18, 33007 Oviedo, Spain}
\affiliation{Instituto Universitario de Ciencias y Tecnolog\'ias Espaciales de Asturias (ICTEA), Calle de la Independencia 13, 33004 Oviedo, Spain}


\begin{abstract}
The surprising manifestation of collectivity in small collision systems, such as nucleon-nucleon and nucleon-nucleus collisions, is perhaps even more striking when discussed at higher momenta. In larger systems, high-$p_T$ elliptic anisotropy is understood as a selection bias effect due to the smaller energy loss experienced along the shorter direction that aligns with the event plane. However, in small systems the amount of energy loss appears insufficient to reproduce the sizable angular anisotropy observed experimentally. In this work, we explore a new mechanism generating preferred orientations for energetic particles without the need of energy loss. We exploit a simple model that is based on two basic although inalienable ingredients: geometry and quantum mechanics. Our findings suggest that this sum-over-paths mechanism can provide a relevant contribution to so-called flow coefficients of energetic particles traversing deconfined media of any size.
\end{abstract}

\maketitle

\section{Introduction}
One of the most surprising properties of the new state of hot QCD matter produced in relativistic heavy-ion collisions  at RHIC and the LHC is its nearly perfect fluidity, characteristic of strongly coupled liquids~\cite{PHENIX:2004vcz,BRAHMS:2004adc,PHOBOS:2004zne,STAR:2005gfr,Gyulassy:2004zy}. In order to efficiently convert spatial anisotropies defined by the initial collision overlap geometry into momentum anisotropies of the soft particles measured at detectors, viscous corrections to the relativistic hydrodynamic equations need to be very small~\cite{Romatschke:2007mq,Song:2010mg,Bernhard:2019bmu}. The intimate relation between the initial geometry of the system and the emergence of preferred orientations is by now well established in a variety of systems~\cite{STAR:2015mki,PHENIX:2018lia}, with hydrodynamics standing as the most plausible link between them. For an approximately elliptic medium (being an ellipse the dominant shape of the overlapping region between the collision of two Lorentz-contracted spherical nuclei), pressure gradients are larger along the event plane axis of the collision than perpendicular to it, and so the subsequent explosion is strongest along the event plane axis. This sets that axis as the preferred orientation for the low-$p_T$ particles resulting from the eventual hadronization of the flowing medium.

The geometry of the overlap region can also be responsible for preferred orientations for high-$p_T$ particles, rare excitations produced within the system which are nevertheless decoupled from it, in the sense that their dynamics is not described by hydrodynamics. Their interactions with the bulk of the system leads to energy loss processes that are part of a set of phenomena colloquially referred to as jet quenching~\cite{dEnterria:2009xfs,Wiedemann:2009sh,Majumder:2010qh,Jacak:2012dx,Muller:2012zq,Connors:2017ptx,Busza:2018rrf,Cao:2020wlm,Cunqueiro:2021wls,Apolinario:2022vzg,Wang:2025lct}. In particular, for an approximately elliptic medium, jets oriented along the semi-minor axis (the event plane of the collision) will typically suffer less energy loss than those oriented along the semi-major axis~\cite{Gyulassy:2000gk,Wang:2000fq}. This leads to a selection bias effect that also sets the event plane axis as the preferred orientation for energetic particles (an elliptic anisotropy), in line with experimental observations in nucleus-nucleus collisions~\cite{STAR:2002pmf,PHENIX:2010nlr,ATLAS:2011ah,CMS:2012tqw,ATLAS:2018ezv,ATLAS:2021ktw,CMS:2022nsv,ATLAS:2024mch}.

While the understanding of the appearance of preferred orientations in the so-called large collision systems (nucleus-nucleus, or $AA$) seems to be more or less complete from low- to high-momentum particles, the situation is very different for small collision systems (nucleon-nucleon, or $pp$, and nucleon-nucleus, or $pA$). The striking discovery of collectivity effects in small systems at the LHC more than a decade ago~\cite{CMS:2010ifv,ATLAS:2012cix,ALICE:2012eyl,CMS:2012qk} (for reviews see~\cite{Nagle:2018nvi,Noronha:2024dtq,Grosse-Oetringhaus:2024bwr}), similar to those observed in large systems, was not accompanied by the distinctive jet quenching phenomena~\cite{ATLAS:2014cpa,ALICE:2016faw,ALICE:2017svf,ALICE:2021wct,ALICE:2023ama}. This is most likely due to the inability of the smallest systems to inflict sizeable energy loss effects onto the traversing energetic particles, as can be inferred by a recent system size scan on the magnitude of high-$p_T$ yield suppression presented in~\cite{CMS:2026qef}. No sizeable quenching effects, alas with a notable exception: significant elliptic anisotropies have been observed in high-$p_T$ particles in $pA$ systems~\cite{ATLAS:2019vcm,ALICE:2022cwa,CMS:2025kzg}. 

The conundrum is immediately clear. How can one obtain a preferred orientation along the event plane axis due to differences in energy loss between the short and the long path, when energy loss effects are too small to start with? This puzzle can be more precisely quantified by noting that in order to obtain the measured high-$p_T$ elliptical anisotropy, also called $v_2$, one needs so much energy loss that yield suppression, also called $R_{AA}$, is way too strong to accommodate experimental observations~\cite{Zhang:2013oca,Roux:2024fpv}.

Naturally, alternative mechanisms to produce high-$p_T$ $v_2$ without sizeable energy loss have been put forward in the literature. These range from correlations in the initial state attributed to gluon saturation~\cite{Dusling:2013oia}, correlations via double parton scattering~\cite{Hagiwara:2017ofm} or independent multiple parton scatterings~\cite{Blok:2017pui,Blok:2018xes}, including as well the role played by momentum conservation of the system as a whole~\cite{Soudi:2023epi,Soudi:2024slz,JETSCAPE:2024dgu}. It has also been observed that in effective kinetic theories of interacting quarks and gluons, a few scatterings are enough to generate sizeable anisotropies via the so-called escape mechanism~\cite{He:2015hfa,Romatschke:2017ejr,Kurkela:2021ctp}. To the best of our knowledge, however, a fully satisfactory explanation is still missing. This motivates us to offer new viewpoints with which to contribute to the resolution of this enticing enigma.

In the last years, there have been important theoretical developments in the computation of jet quenching phenomena which account for non-trivial medium properties such as its inhomogeneity or flowing nature~\cite{Sadofyev:2021ohn,Barata:2022krd,Kuzmin:2023hko,Barata:2023qds,Barata:2024xwy,Barata:2025agq}. In the present work we will take into account another non-trivial, albeit admittedly obvious feature of the medium, which is that it lies within a closed region of space possessing a boundary. In addition to this purely geometrical aspect, the second medium feature we will consider is the following.
It is well known that probe particles under the influence of an external field experience a phase shift. A paradigmatic example is the Aharonov–Bohm effect~\cite{Aharonov:1959fk}, where gauge potentials induce observable phase differences in quantum amplitudes. In gauge theories this phase is encoded in path-ordered exponentials of the gauge field, known as Wilson lines~\cite{Wilson:1974sk}. In jet quenching theory, Wilson lines encode the non-abelian color rotation suffered due to multiple scatterings with the medium (see e.g.~\cite{Wiedemann:2000za,Casalderrey-Solana:2007knd,Mehtar-Tani:2025rty}). The presence of an in-medium phase shift is the key and sole medium-induced effect around which our exploratory model is built. We consider eikonal propagation, without broadening effects nor medium induced radiation, and where the only medium effect consists in a phase shift. It is reasonable to expect that this is the leading effect in small systems, where path lengths are too small for kinematic or inelastic effects to develop. The interplay of the phase shift with the shape of the boundary is the driving force behind the effects presented in this work.

Our model is deliberately simple. It aims at containing just the necessary elements, in its most basic form, appropriate for the physics we want to describe: geometry and quantum mechanics. While this obviously entails the risk of oversimplifying the problem at hand, it is only through this simplicity that the feasibility of the new mechanism can be assessed in this first exploratory study.

The rest of the paper is organized as follows. In Section~\ref{sec:setup} we briefly present our model setup. In Section~\ref{sec:spa} we adopt the stationary phase approximation, appropriate in the high-momentum limit, in order to compute the angular distribution in the far-field where the detector lies, obtaining a transparent expression for $v_2$. Next, in Section~\ref{sec:mat} we compute the angular distribution exactly for the case in which we have an elliptic medium, which allows us to compare against the analytical results of Section~\ref{sec:spa} and assess its regime of validity. In Section~\ref{sec:res} we present our combined results for $v_2$ from Section~\ref{sec:spa} and Section~\ref{sec:mat} as a function of the system size and eccentricity, as well as look at the norm of the transmitted wave, confirming the absence of energy loss effects already for fairly low momenta. In Section~\ref{sec:con} we conclude and look ahead.




\section{Model Setup} \label{sec:setup}
Let us consider a scalar particle $\psi$ that interacts with a constant electrostatic potential $V$ (mostly minus metric convention)
\begin{equation}
    (D^2+m^2)\psi(\bm{r},t)=0 \, ,
\end{equation}
where the covariant derivative is $D_\mu=\partial_\mu+i q A_{\mu}$, with $q A_{\mu}=(V,\bm{0})$.
The time-independent equation that describes the stationary states of energy $E$ is given by the Helmholtz equation
\begin{equation}
\label{eq:helmholtz}
\left(\nabla^2 + k^2\right)\psi(\bm{r}) = 0 \, ,
\end{equation}
where $k$ is the wavenumber, whose value inside and outside the static medium contained in region $\Omega$ is
\begin{align}
k_i = \sqrt{(E-V)^2-m^2}, \quad
k_o = \sqrt{E^2-m^2} \, .
\end{align}
It is important to stress that the effect of the medium leads to no energy loss whatsoever. If the potential $V$ depended on time, the energy $E$ would not be a conserved quantity -- but we have assumed time-independent $V$. The potential $V(\bm{r})$ (which we assume to be piecewise constant in space, namely $V(\bm{r})=V$ if $\bm{r}\in \Omega$ and $V(\bm{r})=0$ if $\bm{r}\notin \Omega$) is real, and therefore the Hamiltonian is Hermitian. Similarly to light entering a glass block, the particle can change its wavenumber $k$, but its frequency $E$ remains fixed. Any reflection and transmission at the boundary will thus be elastic.

This toy model, reminiscent of the century-old $\alpha$-decay problem~\cite{Gamow:1928zz}, despite its simplicity it may capture relevant aspects of high-$p_T$ particle propagation in a medium. The leading effect of the medium on an energetic particle consists in a rotation of its color quantum labels due to multiple soft scatterings with the background gauge field, which are encapsulated in a Wilson line dressing the excitation. If one assumes that the background gauge field is abelian and consists just of a purely electrostatic, time-independent (and space-piecewise-constant) potential, then the Wilson line reduces to $W(t)\sim e^{iVt}$, leading to $\psi(\bm{r},t)=\psi(\bm{r})e^{-i(E-V)t}$. This is clearly equivalent to the modification of the dispersion relation presented above via the minimal coupling prescription.

\section{Large $k$ limit using Stationary Phase Approximation}\label{sec:spa}
The first approach we take is appropriate in the high $k$ limit, and is based in the stationary phase approximation (SPA), or WKB approximation. The goal is to obtain the wavefunction measured at the detector at infinity. Our starting point is the boundary integral representation (a short derivation of which is given in the Supplemental Material), which provides the value of the wavefunction outside of the medium when the source that produced the excitation is inside its boundary
\begin{align}
\label{eq:boundint}    \psi(\bm{r})=\int_{\partial\Omega}\! d\alpha\,\Big[
\psi(\bm s)\,\partial_n G_o(\bm r,\bm s)-G_o(\bm r,\bm s)\,\partial_n\psi(\bm s)
    \Big],
\end{align}
where $\partial \Omega$ represents the boundary of the medium, parametrized by arc-length coordinate $\alpha$ (so that $d\alpha$ denotes the line element along the boundary), $\bm{s}(\alpha)$ is the location of the boundary and $\bm{z}$ is the location of the source. In the far-field approximation, i.e. $|\bm{r}|\rightarrow \infty$, with $\bm{r}=r \hat{\bm{e}}$, the outer Green's function becomes
\begin{align}
    G_o(\bm{r},\bm{s})\sim\frac{ie^{i k_o r}e^{-i\frac{\pi}{4}}}{\sqrt{8\pi k_o r}}e^{-i k_o \bm{s}\cdot \hat{\bm{e}}} \, ,
\end{align}
as can be understood from the asymptotic behavior of $H_0^{(1)}(k_o|\bm{r}-\bm{s}|)$, which is the outgoing wave solution of the Helmholtz equation in two dimensions.
The leading contribution to the boundary integral of the so-called double layer potential term is then $\partial_n G_o(\bm{r},\bm{s})\approx (-i k_o \hat{\bm{e}}\cdot \hat{\bm{n}})G_o(\bm{r},\bm{s})$, with $\hat{\bm{n}}$ the unit vector normal to the boundary. 

We continue by assuming the large wavenumber limit $kR\gg1$, where $R\sim\kappa^{-1}$ is the local curvature radius of $\partial\Omega$, which crucially allows us to use Kirchhoff approximation. For distances close to a given boundary point $\bm{s}$, the wavefront from the incident, small wavelength wave $\psi_i$ can be approximated by a plane wave with local propagation direction $\hat{\bm{u}}\equiv (\bm{s}-\bm{z})/|\bm{s}-\bm{z}|$.
Furthermore, for a wavelength small compared to the local curvature radius of the boundary, the interaction of the plane wave with a smooth interface is well approximated locally by the plane--plane transmission problem, so the total field on the boundary can be written as $\psi(\bm{s},\bm{z})\approx T \,\psi_i(\bm{s},\bm{z})$, where $T$ is the plane-wave transmission amplitude $T=(2k_i\cos{\theta_i})/(k_i\cos \theta_i+k_o\cos\theta_o)$ (see e.g.~\cite{shankar2012principles}). This approximation neglects multiple reflections and other effects that deviate from specular transmission. The angles $\theta_i$ and $\theta_o$ correspond to $\cos\theta_i\equiv \hat{\bm{u}}\cdot\hat{\bm{n}}$ and $\cos\theta_o \equiv \hat{\bm{e}}\cdot\hat{\bm{n}}$. The incident, locally plane, wave created by the point-like source is
\begin{align}
    \psi_i(\bm{s},\bm{z})\sim \frac{ie^{i k_i |\bm{s}-\bm{z}|}e^{-i\frac{\pi}{4}}}{\sqrt{8\pi k_i |\bm{s}-\bm{z}|}} \, ,
\end{align}
and the leading $k$ contribution from the derivative term of the single layer potential term gives $\partial_n \psi(\bm{s},\bm{z})\approx i k_i \cos{\theta_i}\psi (\bm{s},\bm{z})$. Putting everything together, the far-field wavefunction at the detector is
\begin{align}
    \psi(\bm{r})\sim \frac{-e^{i k_o r}}{4 \pi \sqrt{k_o r}} \sqrt{k_i}\int_{\partial \Omega}d\alpha \frac{\cos \theta_i}{\sqrt{|\bm{s}-\bm{z}|}}e^{i \chi(\alpha)} \, ,
\end{align}
where we have defined the highly oscillatory phase as $\chi(s) \equiv k_i |\bm{s}-\bm{z}|-k_o \bm{s}\cdot \hat{\bm{e}}$. To find the stationary points $\alpha^*$, we need to compute the extrema of $\chi(\alpha)$
\begin{equation}
\label{eq:snell}
    \chi'(\alpha)\big|_{\alpha=\alpha^*}=0 \rightarrow k_i \hat{\bm{u}}\cdot \hat{\bm{t}}=k_o \hat{\bm{e}}\cdot \hat{\bm{t}} \, ,
\end{equation}
where we defined the unit tangent vector $\hat{\bm{t}}\equiv d\bm{s}/d\alpha$.
For the second derivative we need to compute
\begin{equation}
\chi''(\alpha) = k_i \frac{d}{d\alpha}(\hat{\bm{u}} \cdot \hat{\bm{t}}) - k_o\frac{d}{d\alpha}(\hat{\bm{e}} \cdot \hat{\bm{t}}) \, .
\end{equation}
The first term yields
\begin{align}
    \frac{d}{d\alpha}(\hat{\bm{u}} \cdot \hat{\bm{t}})=\frac{1}{|\bm{s}-\bm{z}|}\left( 
    \hat{\bm{t}}-(\hat{\bm{u}}\cdot \hat{\bm{t}})\hat{\bm{u}}\right)\cdot \hat{\bm{t}}+\hat{\bm{u}}\cdot (-\kappa \hat{\bm{n}}) \, ,
\end{align}
where $\kappa$ is the local curvature of the boundary and $d\hat{\bm{t}}/d\alpha=-\kappa \hat{\bm{n}}$, with $\hat{\bm{n}}$ the outward normal. Noting that $\hat{\bm{u}}\cdot \hat{\bm{t}}=\sin \theta_i$, we get
\begin{align}
    \chi''(\alpha)=\frac{k_i \cos^2\theta_i}{|\bm{s}-\bm{z}|}+\kappa(k_o \cos\theta_o-k_i\cos{\theta_i}) \, .
\end{align}
Performing the stationary phase approximation then yields
\begin{align}
    \psi(\bm{r}) \sim \frac{-e^{i k_o r}}{4 \pi \sqrt{k_o r}} \frac{\sqrt{k_i}\cos \theta_i}{\sqrt{|\bm{s}^*-\bm{z}|}}\sqrt{\frac{2 \pi}{|\chi''(\alpha^*)|}}e^{i\chi(\alpha^*)+i \frac{\pi}{4} \text{sgn}(\chi''(\alpha^*))} \, ,
\end{align}
where $\bm{s}^*\equiv\bm{s}(\alpha^*)$. From now on we omit overall factors unimportant to the angular dependence at the detector, i.e. we work up to an overall normalization independent of the detector angle. The far-field angular dependent part of the amplitude $|\psi(\bm{r})|^2$ goes like
\begin{align}
\label{eq:dpdphi}
    \mathcal{A}^S(\phi)\sim \frac{k_i \cos^2\theta_i}{|k_i \cos^2\theta_i+L\kappa(k_o \cos\theta_o-k_i\cos\theta_i)|} \, ,
\end{align}
where the superscript $S$ stands for ``SPA''. We have defined the distance from the source to the boundary as $L\equiv|\bm{s}^*-\bm{z}|$. Note that $\theta_i$, $\theta_o$, $L$ and $\kappa$ depend on the location of the stationary point at the boundary, $\bm{s}^*$, which in turn depends on the detector angle $\phi$, as determined by solving the far-field Snell's law in Eq.~\eqref{eq:snell}.


We can get a better handle on the implications of Eq.~\eqref{eq:dpdphi} by making an estimate of $v_2$, the so-called elliptic flow coefficient. To that end we will assume that the source of the particle is at the center of the medium and that the boundary is an ellipse with semi-major axis $a$ and semi-minor axis $b$. If we parametrize the boundary as $x(\theta)=a \cos \theta$ and $y(\theta)=b \sin \theta$, the curvature is
\begin{equation}
    \kappa(\theta)=\frac{ab}{(a^2 \sin^2\theta+b^2 \cos^2 \theta)^{\frac{3}{2}}} \, .
\end{equation}
For a hard probe, an approximate expression for $v_2$ is given by~\cite{Zigic:2018smz}
    \begin{align}
    \label{eq:v2aprox}
        v_2 \approx \frac{1}{2}\frac{\mathcal{A}_{\text{short}}-\mathcal{A}_{\text{long}}}{\mathcal{A}_{\text{short}}+\mathcal{A}_{\text{long}}} \, ,
        \end{align}
        where $\mathcal{A_{\text{short}}}$ ($\mathcal{A_{\text{long}}}$) refers to the intensity in the short (long) direction of the ellipse. Let us take a dispersion relation for a particle with zero mass, i.e. $k_i=k_o-V$, and use 
        the eccentricity $e=\sqrt{1-b^2/a^2}$, finding
        \begin{align}
        \label{eq:v2an}
            v_2^A \approx \frac{V(2-e^2)e^2}{4 (1-e^2)k_o+2e^4V} \, ,
        \end{align}
where the superscript $A$ stands for ``Approximate''. This is one of the main results of this paper. Several interesting observations can be made from this remarkably simple expression. The first two are that when $V=0$ (no medium) or $e=0$ (no spatial anisotropies, as in a circle), then $v_2$ vanishes. One also observes that the size of $v_2$ roughly increases with increasing $e$ and increasing $V$, and it is strictly positive. This expression also predicts that $v_2(k_o)\sim 1/k_o$. Last but not least, this expression for $v_2$ does not depend on the system size.

\begin{figure}[t!]
    \includegraphics[width=0.5\textwidth]{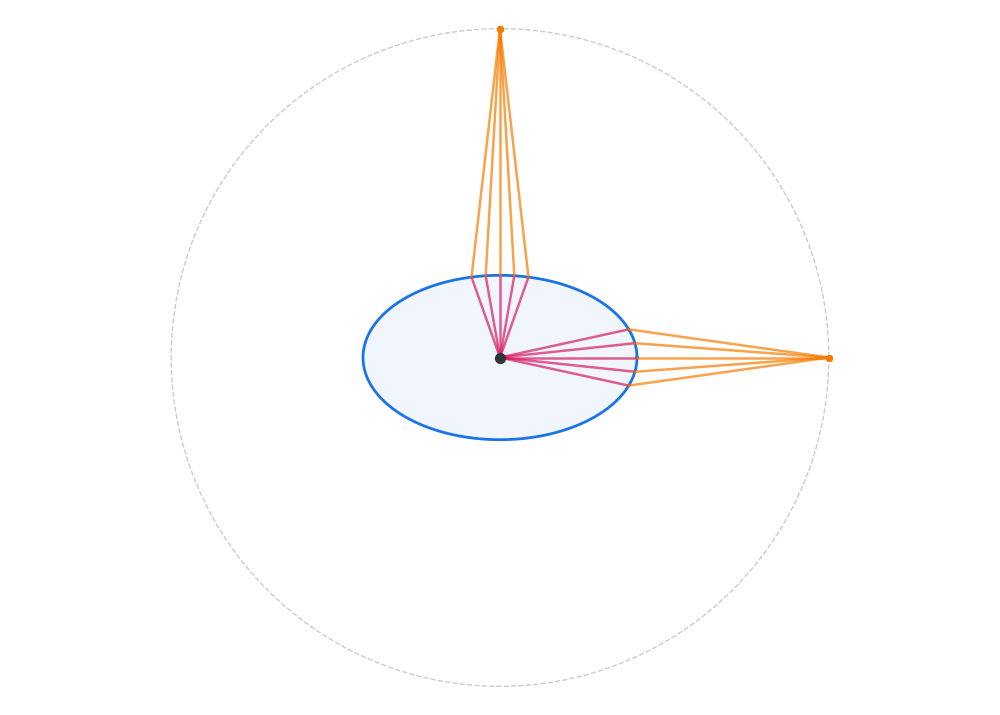}
    \caption{Sum‑over‑paths sketch illustrating the difference in in-medium phase accumulation between trajectories exiting along the short (lower curvature $\kappa$) and long (higher curvature $\kappa$) directions of the ellipse. The amount of paths explored is proportional to the particle wavelength $\lambda$. Sizes are not drawn to scale: the medium (in blue) is only a few femtometers across, whereas the detector (dashed grey line) is located meters away.}
    \label{fig:Idea_Caminos}
\end{figure}

        The physics of this result can be easily understood from the sum-over-paths picture of quantum mechanics. A particle going along the short direction will encounter a smaller curvature $\kappa$ at the boundary than if it goes along the long direction. At small curvatures $\kappa$, different paths nearby the stationary point (those explored by the particle wavelength $\lambda$), will acquire an in-medium phase that is relatively similar. When curvature is high, the phases of those nearby paths become increasingly different. This logic is depicted with the help of Fig.~\ref{fig:Idea_Caminos}. The result of having similar phases is that they add more constructively than when they are disparate, thus enhancing the amplitude in the short direction (low curvature) more than in the long direction (high curvature). How fast the phase varies is encoded in the second derivative term stemming from the stationary phase approximation, $|\Chi''(\alpha^*)|$. As the particle wavelength decreases, the size of the boundary explored also decreases, making differences in in-medium paths between the low and high curvature exit points fade away as the boundary becomes locally flat. This last observation explains why the size of $v_2$ decreases with increasing $k\sim 1/\lambda$.


The analytical expression in Eq.~\eqref{eq:v2an} is useful to discuss the qualitative behavior of $v_2$; however, for the highest eccentricities the actual value can sizeably deviate from the approximate one. For this reason, in Section~\ref{sec:res} we show results for the full SPA expression in Eq.~\eqref{eq:dpdphi}. Details on how it is solved are given in the Supplemental Material.

\section{Exact solution for an elliptic medium using Mathieu functions}\label{sec:mat}
In order to obtain a solution that does not rely on the type of approximations adopted in the SPA analysis, we now solve the problem numerically. In particular, this approach allows us to explore arbitrary values of $k_o$, such as those which are not necessarily larger than $1/R$. From the beginning, we restrict the shape of the medium to be that of an ellipse. 
The natural coordinates are standard elliptic coordinates $(\mu, \nu)$
\begin{align}
x &= c \cosh\mu \cos\nu \, ,\nonumber \\ 
y &= c \sinh\mu \sin\nu \, ,
\end{align}
with $\mu \geq 0$ and $0 \leq \nu < 2\pi$. The curves of constant $\mu$ are ellipses with a major semi-axis $a = c \cosh\mu$
and a minor semi-axis $b = c \sinh\mu$, with an eccentricity 
$e =1/\cosh\mu_b$, where $\mu_b$ denotes the boundary of the elliptic medium. In elliptic coordinates, the Laplacian is given by $\nabla^2 = \left( \partial_{\mu}^2 + \partial_{\nu}^2 \right)/\sqrt{g}$, with $\sqrt{g}=c^2(\cosh^2\mu-\cos^2\nu)$. Using the separation Ansatz $\psi(\mu, \nu) = M(\mu) N(\nu)$, the Helmholtz equation in Eq.~\eqref{eq:helmholtz} becomes
\begin{equation}
\frac{1}{M(\mu)} \frac{d^2 M}{d\mu^2} + \frac{1}{N(\nu)} \frac{d^2 N}{d\nu^2} + 2q(\cosh 2\mu - \cos 2\nu) = 0 \, ,
\end{equation}
where we have defined the dimensionless parameter $q=k^2 c^2/4$.
This separates into two ordinary differential equations -- the angular (in $\nu$) and radial (in $\mu$) Mathieu equations
\begin{align}
\frac{d^2 N}{d\nu^2} &+ (\lambda_q - 2q \cos 2\nu) N = 0 \, , \nonumber \\
\frac{d^2 M}{d\mu^2} &- (\lambda_q - 2q \cosh 2\mu) M = 0 \, ,
\end{align}
where $\lambda_q$ is the separation constant. Periodic solutions in $\nu$ exist only for a discrete set of eigenvalues, $\lambda_q^m$, which satisfy the angular Mathieu equation with even or odd parity, where the index $m$ labels the modes (representing the number of nodes over the ellipse). The Fourier coefficients for the even and odd angular solutions, $\operatorname{ce}_m(\nu,q)$ and $\operatorname{se}_m(\nu,q)$, are determined by substituting a Fourier expansion into the angular equation and solving the resulting recurrence relations. The corresponding radial solutions, $\operatorname{Mc}_m(\mu,q)$ and $\operatorname{Ms}_m(\mu,q)$, are obtained using the same set of eigenvalues $\lambda_q^m$ entering the angular equation, but are expressed as a Bessel series instead of a Fourier series due to the transformation $\nu \rightarrow i \mu$. 

Using the addition theorem~\cite{abramowitz1964handbook}, we can express the Green's function of the two-dimensional Helmholtz equation, $G_o(\bm{r})=\tfrac{i}{4}H_0^{(1)}(k r)$, in terms of Mathieu functions in elliptic coordinates as
\begin{align}
    &H_0^{(1)}(kr) = \frac{2}{\pi} \sum_{m=0}^\infty \Big[ \operatorname{ce}_m(\nu, q) \operatorname{ce}_m(\nu_s, q) \operatorname{Mc}_m^{(1)}(\mu_s, q) \times \nonumber\\&\operatorname{Mc}_m^{(3)}(\mu, q)
+ \operatorname{se}_m(\nu, q) \operatorname{se}_m(\nu_s, q) \operatorname{Ms}_m^{(1)}(\mu_s, q) \operatorname{Ms}_m^{(3)}(\mu, q) \Big]
\end{align}
where we are covering the region $\mu>\mu_s$, with $(\mu_s,\nu_s)$ the location of the point-like source within the ellipse. The radial Mathieu function of the first kind $\operatorname{Mc}_m^{(1)}$ is regular at the focal points, and represents standing waves (the elliptic counterpart of the Bessel functions). The radial function of the third kind, $\operatorname{Mc}_m^{(3)}=\operatorname{Mc}_m^{(1)}+i \operatorname{Mc}_m^{(2)}$, represents traveling waves (analogously to Hankel functions), with the second kind radial function, $\operatorname{Mc}_m^{(2)}$, blowing up at the focal points (analogously to Neumann functions).

From now on we write only the even sector; the odd sector is obtained by the substitutions $\operatorname{Mc}\rightarrow \operatorname{Ms}$ and $\operatorname{ce}\rightarrow \operatorname{se}$. The equations that determine the transmission and reflection coefficients decouple between the even and odd sectors. Note that the sum over modes $m$ always starts with $m=1$ for the odd sector.

Our goal is to obtain the wavefunction outside of the medium, where $\mu>\mu_b$, so we are mainly concerned with the outer solution
\begin{align}
\psi_{o}(\mu, \nu) = \sum_{m=0}^\infty C_m \operatorname{ce}_m(\nu, q_o) \operatorname{Mc}_m^{(3)}(\mu, q_o)
+ (\text{odd})\, .
\end{align}
The even and odd coefficients, $C_m$ and $D_m$, will need to be determined by matching the inner and outer solutions at the boundary, $\psi_i(\mu_b,\nu)=\psi_o(\mu_b,\nu)$, along with their derivatives with respect to the normal at the boundary, $\partial_\mu \psi_i(\mu_b,\nu)=\partial_\mu\psi_o(\mu_b,\nu)$. The great convenience of using elliptical coordinates for our problem is now evident -- one of the coordinates describes precisely the boundary of the medium. The wavefunction inside the ellipse, with $\mu < \mu_b$, is $\psi_i(\mu, \nu) = \psi_s(\mu, \nu) + \psi_r(\mu, \nu)$, where $\psi_s$ and $\psi_r$ are the source wave, or incident wave, and the reflected wave, respectively. Accordingly, we expand them as
\begin{align}
\psi_s = \sum_{m=0}^\infty &\alpha_m \operatorname{ce}_m(\nu, q_i) \operatorname{Mc}_m^{(3)}(\mu, q_i) 
+ (\text{odd}) \, , \\ 
\psi_{r} = \sum_{m=0}^\infty & A_m \operatorname{ce}_m(\nu, q_i) \operatorname{Mc}_m^{(1)}(\mu, q_i) 
+ (\text{odd}) \, , 
\end{align}
where the source wave is traveling from the source to the boundary and the reflected wave is required to be regular at the focal points as it bounces back from the boundary.

\begin{figure*}[t]
    \centering
    \includegraphics[width=1\textwidth]{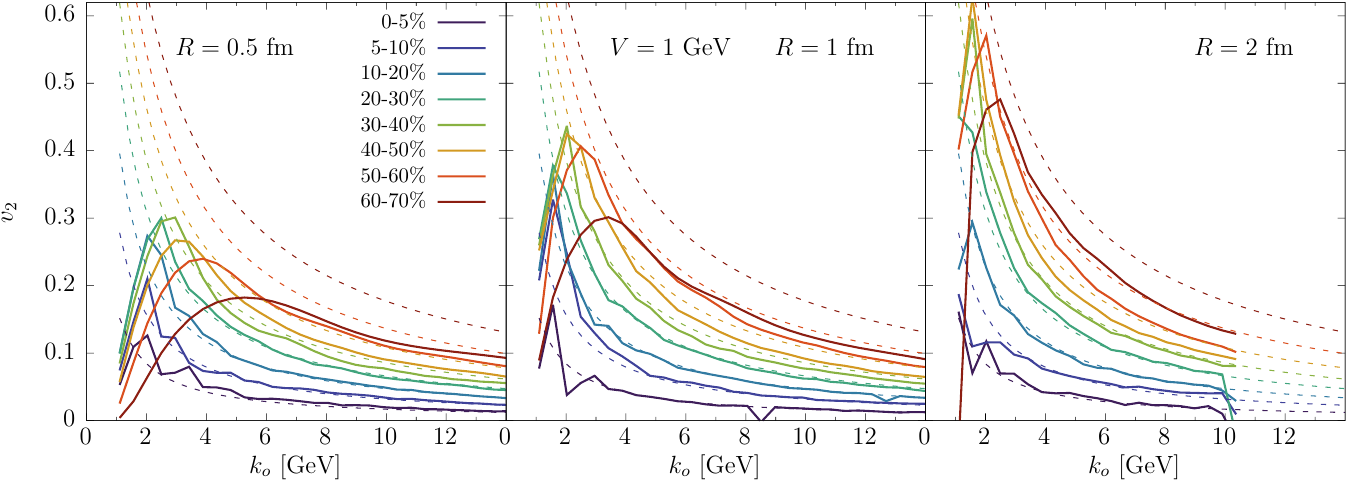}
    \caption{Results for $v_2$ versus momentum $k_o>V=1$ GeV using the event plane method, as a function of centrality, for three different ``nucleus'' sizes: $R=0.5$ fm in the left, $R=1$ fm in the center and $R=2$ fm in the right. Dashed lines correspond to the SPA solution $v_2^S$ (identical in all three panels) while solid lines show the exact solution using Mathieu functions $v_2^M$.}
    \label{fig:pot1vsan}
\end{figure*}

For a general source distribution $S(\mu',\nu')$ of the excitation, the incident field can be determined as
\begin{multline}
    \psi_s(\mu,\nu)
    =\int d\mu'\,d\nu'\,\sqrt{g}\,
    G_i\!\left(\mu,\nu;\mu',\nu'\right)\,
    S(\mu',\nu') \, ,
\end{multline}
with $G_i$ the Green's function inside the medium. 
In particular, for a point-like source at $(\mu_s,\nu_s)$ one has
$S(\mu',\nu')=\delta(\mu'-\mu_s)\,\delta(\nu'-\nu_s)/\sqrt{g(\mu_s,\nu_s)}$, 
so that $\psi_s(\mu,\nu)=G_i(\mu,\nu;\mu_s,\nu_s)$ trivially. Inserting the Mathieu expansion of $G_i$ one readily finds the (even) incident wavefunction coefficients
\begin{align}
\alpha_m &= \frac{i}{2\pi} \operatorname{ce}_m(\nu_s, q_i) \operatorname{Mc}_m^{(1)}(\mu_s, q_i)\, .
\end{align}
Having determined the source, or incident wavefunction $\psi_s$, we can compute the $C_m$ and $D_m$ coefficients of $\psi_o$, which are related to the transmission coefficient and will provide the wavefunction for the detector in the far-field. Imposing the matching conditions at the boundary, the resulting equations for the (even) coefficients are
\begin{align}
\label{eq:lineqs}
    &\displaystyle\sum_{m\geq 0} O_{nm}^{c}(q_i,q_o)\mathcal{W}_{mn}^{c}(q_o,q_i,\mu_b)C_m = 
    i\pi\alpha_n \, ,
\end{align}
with (even) ``Wronskians'' defined as
\begin{align}
\mathcal{W}_{mn}^{c}(q_o,q_i,\mu_b) = &\operatorname{Mc}_m^{(3)\prime}(\mu_b, q_o) \operatorname{Mc}_n^{(1)}(\mu_b, q_i) 
- \nonumber \\
&\operatorname{Mc}_m^{(3)}(\mu_b, q_o) \operatorname{Mc}_n^{(1)\prime}(\mu_b, q_i) \, ,
\end{align}
and we have used that $\mathcal{W}_{mm}^{c}(q,q,\mu)=i$.
$O_{mn}^{c}$ is the (even) overlap matrix that arises from the fact that the angular Mathieu functions are not orthogonal between different $q$ values:
\begin{align}
    O^{c}_{nm} = \displaystyle\int_0^{2\pi}\operatorname{ce}_n(\nu,q_i)\operatorname{ce}_m(\nu,q_o)d\nu \, .
\end{align}
When $q_i=q_o$ (no medium) the overlap matrix is diagonal and Eqs.~\eqref{eq:lineqs} yield $C_m= \alpha_m$, this is, $\psi_o=\psi_s$. Note that to obtain Eqs.~\eqref{eq:lineqs} we have used two equations relating $A_m$ with $C_m$. The problem of determining the outer wavefunction is thus reduced to simply inverting a linear operator. The transmission coefficient for the odd modes can be solved by the same procedure.



We are interested in the behavior of the wave function at the detector at infinity, which, in elliptic coordinates, corresponds to taking $\mu \to \infty$; in that limit, $\nu$ plays the role of the angle $\phi$ in usual polar coordinates. The radial Mathieu wavefunctions take the asymptotic form
\begin{align}
\label{eq:asymptotic_Ms3}
    (\operatorname{Mc,Ms})_m^{(3)}(\mu, q_o)\sim \sqrt{\frac{2}{\pi k_o \rho}} \exp\left[i\left(k_o \rho - \frac{m\pi}{2} - \frac{\pi}{4}\right)\right] \, ,
\end{align}
where $\rho \approx \tfrac{a}{2} e^\mu$. Up to unimportant overall factors for the angular dependence of the wavefunction, we have
\begin{align}
    \psi_o(\phi) \sim
        \sum_{m=0}^\infty \tilde{C}_m \operatorname{ce}_m(\phi, q_o)
        + (\text{odd}) \, ,
\end{align}
where we have defined $\tilde{C}_m \equiv e^{-im\pi/2}C_m$. 
Finally, the angular-dependent amplitude is
\begin{eqnarray}
\label{eq:matamp}
    \mathcal{A}^M(\phi) \sim |\psi_o|^2 \, ,
\end{eqnarray}
where the superscript $M$ stands for ``Mathieu''.
\section{Results}\label{sec:res}
The observable we are interested in is the elliptic flow coefficient $v_2$. Both for the case of the exact solution using Mathieu functions and the SPA solution, we adopt the event plane method
\begin{equation}\label{v2}
    v_2^{M,S} \equiv \frac{\displaystyle\int_{0}^{2\pi} d\phi \, \cos(2(\phi-\Psi_2))\, \mathcal{A}^{M,S}(\phi)}
    {\displaystyle\int_{0}^{2\pi} d\phi \, \mathcal{A}^{M,S}(\phi)} \, ,
\end{equation}
where $\Psi_2=\pi/2$, with $\mathcal{A}^S(\phi)$ defined in Eq.~\eqref{eq:dpdphi} and $\mathcal{A}^M(\phi)$ in Eq.~\eqref{eq:matamp}.

In order to study the dependence of $v_2$ on the system size $a$ and eccentricity $e$, or in the focal distance $c=a e$, let us constrain the geometric parameter space by considering a situation that echoes the so-called centrality evolution in heavy-ion collisions. For two circles of radius $R$ with impact parameter $I$, we can approximate the overlapping region as an ellipse. The semi-minor axis would be $b=R-I/2$, with a semi-major axis $a=\sqrt{R^2-I^2/4}$. The focal distance squared is then $c^2=a^2 e^2=a^2-b^2=IR-I^2/2$. For a minimum impact parameter of $I=0$ and a maximum of $I=2R$, we see that $c^2$ starts from 0, attains a maximum value at $I=R$ and then goes to 0 again. The eccentricity is $e=\sqrt{(4IR-2I^2)/(4R^2-I^2)}$, and goes from $e(I=0)=0$ to $e(I=2R)\rightarrow 1$. In a heavy-ion collision, the 10\% of all events with the highest particle multiplicity belong to the 0-10\% centrality class, the next 10\% of events with the second-lowest belong to the 10-20\% class, and so on. Since multiplicity scales monotonically with $I$ (lower $I$, more multiplicity), the cumulative fractions in multiplicity map to those in impact parameter $I$. The ranges for $I$ we take are $I=R\lbrace 0,0.5,0.7,1,1.2,1.4,1.6,1.7,1.9\rbrace$, e.g. from Glauber model simulations for PbPb collisions~\cite{ALICE:2013centrality}, which as we argued approximately follow the geometric distribution for $I$, namely $I dI$. We just show results for the average $I$ in a given centrality bin, namely $\langle I \rangle =(2/3)(I_M^3-I_m^3)/(I_M^2-I_m^2)$, where $I_M$ and $I_m$ are the maximum and minimum impact parameters of that class. For a given ``nucleus" size $R$, the average $I$ of a centrality class thus determines both $c(\langle I\rangle)$ and $e(\langle I \rangle)$\footnote{Since it does not make sense to select $pA$ collisions in terms of impact parameter, the centrality classes used here should not be interpreted literally, and should rather be thought as ways of selecting configurations with similar shapes or eccentricities.}.

Results are shown in Fig.~\ref{fig:pot1vsan}. $v_2^M$ is depicted with solid lines, while $v_2^{S}$ is in dashed lines. The three panels correspond to different sizes: $R=0.5$ fm in the left, $R=1$ fm (the proton radius) in the center and $R=2$ fm in the right. All results are obtained for a massless excitation with $V=1$ GeV. This value of the potential has been chosen so as to obtain a $v_2\sim 0.1$ at around $k_o\sim 10$ GeV for the 30-40\% centrality class. The mode sums over $m$ for the exact solution with Mathieu functions needs to be truncated at some value $m_{\rm max}$, which roughly scales like $m_{\rm max}\sim 2 \sqrt{\max{(q_i,q_o)}}$. We have cut the $v_2^M$ solution at $k_o=10$ GeV for ``nucleus'' size $R=2$ fm. The reason is that the numerical implementations of the radial Mathieu functions we use, those in \texttt{GSL}, start to fail at very large $q\sim R^2 k_o^2/4 \sim 2500$ (one way in which they are seen to fail is that the computed $v_2$ does not vanish when $V=0$). Getting to arbitrarily large values of $R$ and $k_o$ will require switching to a different numerical implementation or using asymptotic expressions.

The first thing we observe is that the SPA solution $v_2^{S}$ (dashed) captures quite well the trend observed in the exact $v_2^M$ results (solid), especially for lower eccentricities (more central collisions) where the agreement is quantitative down to fairly low $k_o$. We have to add that the $v_2^M$ solutions correspond to the average over many source positions, $\mathcal{O}(70)$\footnote{It is a norm-weighted average, since for each value of $k_o$, contributions from different source positions can in principle be different. There is very little difference with respect to an unweighted average.}, while the $v_2^{S}$ is just for a source at the center (there are only small differences if one averages the SPA solution over source positions, as can be seen in the Supplemental Material). The fact that $v_2^M$ agrees so well with $v_2^S$ for all three panels, where the system size is varied up to a factor four, is a confirmation that the mechanism produces a high-$k_o$ $v_2$ whose magnitude depends only on the eccentricity and the potential strength, but not on the medium size.

For many of the curves shown in Fig.~\ref{fig:pot1vsan}, as we decrease $k_o$, $v_2^M$ peaks, and then drops. We can understand this behaviour in terms of the size of the wavelength of the excitation compared to the medium size. As the wavelength becomes larger, the shape of the boundary is blurred, and the excitation no longer resolves the features that cause the angular anisotropy. Since more peripheral centrality classes (e.g. 50-60\%) are actually smaller in size than the more central ones (e.g. 5-10\%), for a given size $R$ (a given panel), the drop starts at a smaller wavelength, namely at a larger $k_o$. This is what we observe in regards to the centrality dependence of the peak position for all three panels. This also explains why the SPA solution agrees better with the exact solution for more peripheral centrality classes, and down to lower values of $k_o$, in the right panel (larger size) than in the left panel (smaller size) -- the assumption that $k_o R \gg 1$ is more valid the larger the size.
\begin{figure}[t]
    \centering
    \includegraphics[width=0.47\textwidth]{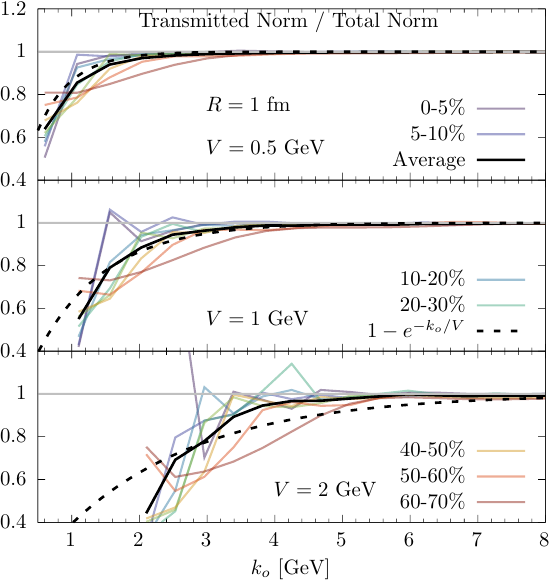}
    \caption{Ratio of the transmitted wave norm (with $V\neq0$) divided by the source norm (or setting $V=0$) versus momentum $k_o>V$, as a function of centrality, for different values of the potential $V$ and fixed ``nucleus'' size $R=1$ fm. The solid black line represents the centrality-averaged result, and the dashed line simply shows the curve for $1-e^{-k_o/V}$.}
    \label{fig:normvspot}
\end{figure}

Note that we have chosen to show results for momentum values $k_o>V$. The regime where $k_o<V$, where the probability current associated to the transmitted wave is very strongly modified, will be discussed in future work. 
Indeed, for values of $k_o$ approaching $V$ (even from above, as studied presently), the norm of the transmitted wave is no longer the same as if there was no medium. We can quantify the probability that the excitation gets ``trapped" within the medium by plotting the ratio of the norm, $\oint d\phi \mathcal{A}^M(\phi)$, when $V\neq 0$ (in medium) by the norm when $V=0$ (in vacuum), as shown in Fig.~\ref{fig:normvspot} for different values of the potential and fixed ``nucleus'' size $R=1$ fm. The top panel has $V=0.5$ GeV, the middle panel $V=1$ GeV and the bottom panel $V=2$ GeV. We observe that the ratio rapidly goes to one already for relatively low values of $k_o$. This means that even though the probability flux has been angularly redistributed according to the medium geometry, essentially none belongs to the reflected wave; the excitation has effectively escaped the medium. Indeed, even when the even and odd reflected coefficients, $A_m$ and $B_m$, go to zero, the transmitted coefficients, $C_m$ and $D_m$, maintain a non-trivial relation to the source wavefunction coefficients, $\alpha_m$ and $\beta_m$, because of the non-orthonormal mode matching at the boundary due to $q_i\neq q_o$. It is interesting to note that the centrality-averaged result closely follows the curve $1-e^{-k_o/V}$, which means that the norm ratio associated to the reflected wavefunction is $e^{-k_o/V}$. The fate of the probability flux associated to the reflected wavefunction, together with the momentum regime $k_o<V$, will be studied in future work.

To conclude, the results shown in Fig.~\ref{fig:normvspot} quantify the momentum above which we can say that there is no spectrum shift due to probability loss in our setup, implying nuclear modification factors (such as the so-called $R_{AA}$) that are consistent with one, i.e. no yield modification. This is why one can say that within our model one obtains sizeable $v_2$ without ``energy loss'' for high-momentum particles.

\section{Conclusions and Outlook}\label{sec:con}
In this work we have presented a new mechanism with which to generate sizeable elliptic anisotropies for energetic particles without the need of energy loss. These results are of particular phenomenological relevance given the striking magnitude of elliptic anisotropy measured for high-$p_T$ particles in small systems, such as $pA$~\cite{ATLAS:2019vcm,ALICE:2022cwa,CMS:2025kzg}, where energy loss effects seem to be too small to be measured. These experimental observations have long been regarded as a puzzle, challenging our understanding of the origin of collectivity in small systems. 

The physics behind the origin of elliptic flow $v_2$ in our model are rooted in the interplay between the quantum mechanical nature of particles and the shape of the boundary of the medium in which they are produced. The only medium effect considered on the particles consists in a phase shift, which we argue is the leading effect in small systems where path lengths are not large enough to trigger medium-induced radiation or transverse momentum broadening. Different exit points in the medium correspond to different local boundary properties, such as different local curvature. At lower curvature regions the explored nearby paths (how nearby depends on the particle wavelength) will have more similar phase shifts than at higher curvature regions, resulting in a less destructive interference. This translates into preferred orientations where a particle is more likely to exit through different points at the boundary than others.

We have provided transparent analytical expressions and have checked their regime of validity using exact numerical solutions for the specific case of an elliptic medium. We have found that the magnitude of the elliptic anisotropy $v_2$ is positive, scales with the eccentricity and the medium potential strength, decreases with increasing particle wavelength and is independent of the medium size (it is thus a conformal, scale invariant effect). By looking at the norm of the transmitted wavefunction, the only contribution studied in this work, we conclude that the particle exits the medium with probability close to one (i.e. no ``energy loss'', or spectra shifts due to medium effects) already at fairly low energies. Crucially, then, although energetic particles escape the medium intact, the escape probability is anisotropic, leading to a large $v_2$.

While these results are encouraging, the model used to describe them is essentially a toy model, containing just the simplest versions of the few key necessary ingredients. An exhaustive list of all the ways this exploratory model can be improved would be too long to enumerate, but there are a few points we can mention. To start, a natural first step would be to consider actual non-abelian Wilson lines in QFT rather than a constant electrostatic potential in quantum mechanics.
More realistic medium properties should also be added, such as medium boundary shapes determined by the fluctuating nature of subnucleonic degrees of freedom~\cite{STAR:2025ivi}, or spacetime- and/or energy-density-dependent background gauge fields, which can even be stochastic. (On this note, the assumption that we have a constant potential $V$ over the length of the (small) medium resembles the situation explored in~\cite{Aurenche:2012qk} within the Color Glass Condensate model, as if we had a single color domain). Introducing geometry and energy density fluctuations (or even quantum superposition effects in the shape itself~\cite{Ke:2025tyv}) would allow, among others things, to obtain sizeable values for triangular anisotropies as well~\cite{Alver:2010gr}, all while reducing the magnitude of $v_2$, which, naturally, is maximized for an exact ellipse.

It is worth mentioning that even though there is a reasonable phenomenological description of elliptic anisotropies of energetic particles in large systems, achieving a quantitative description (without too much energy loss) has since long been regarded as a standing theoretical challenge~\cite{Shuryak:2001me,Liao:2008dk,Betz:2014cza,Noronha-Hostler:2016eow,Andres:2019eus,Stojku:2020wkh,Huss:2020whe,Zhao:2021vmu,Pablos:2025cli}. Given that the sum-over-paths mechanism presented in this paper is independent of the system size, one should expect a contribution to the $v_2$ of energetic particles in large systems as well, in addition to that coming from energy loss or coalescence effects~\cite{Molnar:2003ff}. It will also be very interesting to see upcoming experimental results on the $p_T$-dependence of $v_2$ for OO and NeNe collisions (ATLAS measured it but only up to 6 GeV~\cite{ATLAS:2025nnt}), which are systems that fill the gap between large and small systems, with a nuclear radius of $R\approx 3$ fm. Even though sizeable energy loss effects have been measured~\cite{CMS:2025bta,ATLAS:2025ooe}, to what extent these alone will be able to describe elliptic anisotropies at high-$p_T$ can shed light on the relative importance of alternative mechanisms such as the one here presented.

\section*{Acknowledgements}
We are grateful to Jorge Casalderrey-Solana, Carlos Hoyos, Patrick Meessen, Daniele Musso and Andrey Sadofyev for helpful conversations related to this project. EC and DP are supported in part by the Spanish national grant MCIU-22-PID2021-123021NB-I00. DP is supported by the Ram\'on y Cajal fellowship RYC2023-044989-I.

\bibliographystyle{apsrev4-1}
\bibliography{main.bib}

 \begin{widetext}
%

\section*{Supplemental Material}
\setcounter{equation}{0}
\renewcommand{\theequation}{S\arabic{equation}}

\subsection{Proof of boundary integral representation formula}
Here we present the derivation of Eq.~\eqref{eq:boundint} in the main text. Let $\Sigma\subset\mathbb{R}^2$ be a domain with a smooth boundary $\partial\Sigma$ and outward unit normal $\hat{\bm n}$.
Consider two sufficiently smooth functions $u(\bm x)$ and $v(\bm x)$. Green's second identity states that
\begin{equation}
\label{eq:green}
    \int_{\Sigma}\! d^2x\,\Big(u\,\nabla^2 v - v\,\nabla^2 u\Big)
    = \int_{\partial\Sigma}\! d\alpha\,\Big(u\,\partial_n v - v\,\partial_n u\Big),
\end{equation}
where $\partial_n \equiv \hat{\bm n}\cdot\nabla$ and $d\alpha$ is the line element along the boundary.
We apply this identity to the Helmholtz equation shown in Eq.~\eqref{eq:helmholtz}. Let $\psi$ be a solution of
\begin{equation}
    \left(\nabla^2 + k_o^2\right)\psi(\bm x)=0
\end{equation}
in the exterior region (outside the medium, where there are no sources), defined by $\Sigma$, and let $G_o(\bm x,\bm r)$ be the corresponding outer Green's function, defined by
\begin{equation}
    \left(\nabla^2 + k_o^2\right)G_o(\bm x,\bm r)=-\delta^{(2)}(\bm x-\bm r),
\end{equation}
with the Sommerfeld radiation condition at infinity.
If we choose $u(\bm x)=\psi(\bm x)$ and $v(\bm x)=G_o(\bm x,\bm r)$, then the left hand side of Eq.~\eqref{eq:green} is just $-\psi(\bm{r})$, with $\bm{r}\in \Sigma$.
The integral over $\partial \Sigma$ in the right hand side of Eq.~\eqref{eq:green}, which is the boundary of the exterior region, can be separated into two parts:
the boundary of the medium, $\partial \Omega$, and a boundary with arbitrary shape far away from the medium, $C$, so $\partial \Sigma = \partial \Omega \cup C$. If we take the limit where we place $C$ at infinity, the fields vanish at that boundary and we are just left with the contribution at $\partial \Omega$. Since at boundary $\partial \Omega$ the normal $\hat{\bm{n}}$ points into the medium as viewed from the exterior region, we redefine it such that it now points outwards the medium, i.e. $\hat{\bm{n}}\rightarrow -\hat{\bm{n}}$, flipping the sign of the right hand side as a result. By equating the left and right hand sides of Eq.~\eqref{eq:green}, we obtain the solution for the wavefunction outside the medium in terms of the Dirichlet and Neumann data at the boundary as

\begin{equation}
    \psi(\bm r)
    = \int_{\partial\Omega}\! d\alpha\,\Big[
    \psi(\bm s)\,\partial_n G_o(\bm r,\bm s)-G_o(\bm r,\bm s)\,\partial_n\psi(\bm s)
    \Big] \, ,
\end{equation}
with $\bm{s}\in \partial \Omega$. This is the boundary integral representation in Eq.~\eqref{eq:boundint} and the starting point of the stationary phase approximation analysis of Section~\ref{sec:spa}.

\subsection{Finding stationary points for an elliptic medium}
\label{sec:snell_supp}
In this Section we outline how we determine the boundary stationary points $\bm{s}^*$ that extremize the rapidly oscillating phase $\Chi(s)$. This is necessary to compute the angular dependent amplitude $\mathcal{A}^S(\phi)$ in the stationary phase approximation, which then determines $v_2^S$.

\begin{figure}[t!]
    \centering
    \includegraphics[width=0.7\linewidth]{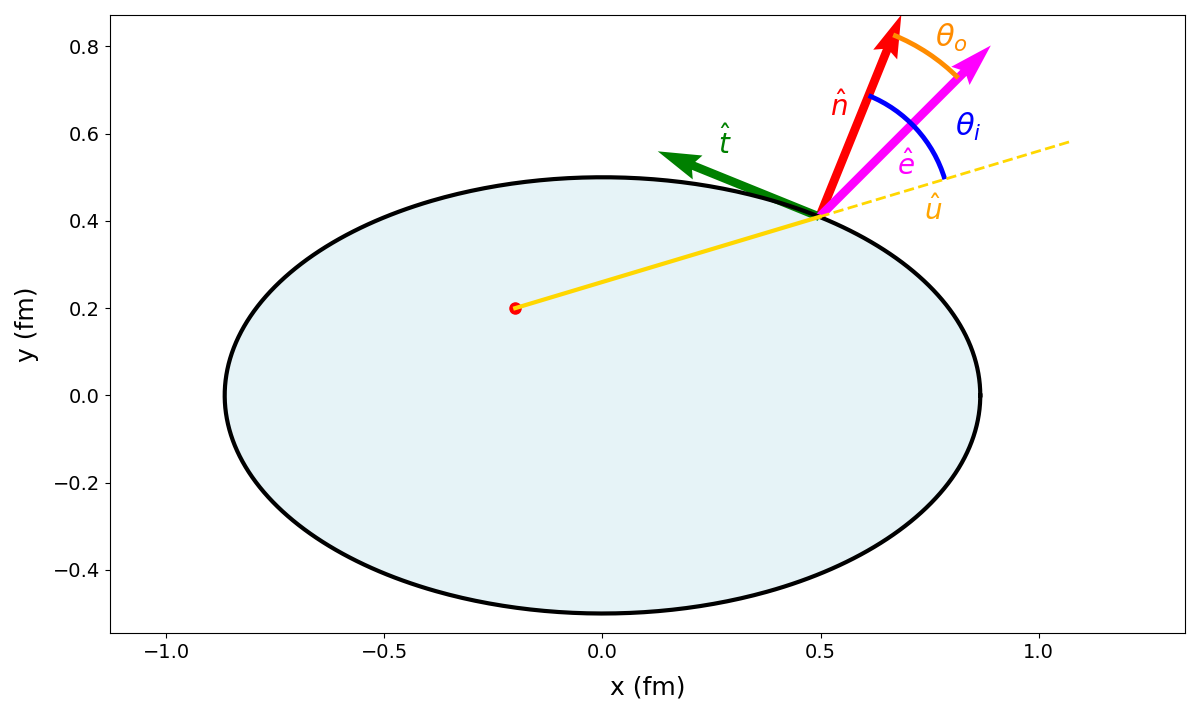}
\caption{Example of a specific configuration
depicting the relevant quantities involved in the determination of the stationary points via Eq.~\eqref{eq:snellnum}. The detector angle is $\phi=\pi/4$, with $k_o=2$ GeV and $V=1$ GeV. The source location, $\bm{z}$, is at the red dot. The particle trajectory inside the medium, $\hat{\bm{u}}$, is shown in yellow, and the one outside, $\hat{\bm{e}}$, in purple. The tangent vector to the trajectory ($\hat{\bm{t}}$, green), the ellipse normal ($\hat{\bm{n}}$, red), the incident angle $\theta_i$ (blue), and the outgoing angle $\theta_o$ (orange) are drawn at the boundary exit point satisfying the stationary solution condition, $\bm{s}^*$.}
    \label{fig:ex_snell}
\end{figure}

The equation that one needs to solve is Eq.~\eqref{eq:snell}
\begin{equation}
\label{eq:snellapp}
k_i \hat{\bm{u}}\cdot \hat{\bm{t}}=k_o \hat{\bm{e}}\cdot \hat{\bm{t}} \, ,
\end{equation}
namely Snell's law in the far-field approximation. For each detector angle $\phi$, entering $\hat{\bm{e}}=(\cos \phi,\sin \phi)$, we want to find the point at the boundary that satisfies Eq.~\eqref{eq:snellapp}. If we parametrize the boundary as an ellipse with coordinates $\bm{s}(\theta)=(a \cos \theta,b\sin \theta)$, with $a$ and $b$ the semi-major and semi-minor axis, respectively, then the tangent unit vector is $\hat{\bm{t}}(\theta)=(-a \sin \theta,b \cos \theta)/\sqrt{a^2+b^2}$, and the unit vector oriented from the source location to the boundary is $\hat{\bm{u}}=(a \cos\theta-x_s,b \sin \theta-y_s)/\sqrt{(a \cos\theta-x_s)^2+(b \sin \theta-y_s)^2}$, with source location $\bm{z}=(x_s,y_s)$. If we plug these into Eq.~\eqref{eq:snellapp} we get
\begin{equation}
\label{eq:snellnum}
    \frac{k_i}{\sqrt{(a\cos\theta^*-x_s)^2+(b\sin\theta^*-y_s)^2}}(-a\sin\theta^*(a\cos\theta^*-x_s)+b\cos\theta^*(b\sin\theta^*-y_s)) =
    k_o(-a\sin\theta^*\cos\phi+b\cos\theta^*\sin\phi).
\end{equation}
Then, for given $k_i$ and $k_o$ (with $k_i$ related to $k_o$ via the potential strength $V$, assuming a massless excitation), a given source location $(x_s,y_s)$, and a given detector angle $\phi$, we find the angle at the boundary $\theta^*$ for which Eq.~\eqref{eq:snellnum} is satisfied. We do this numerically, keeping only the ``bright path'', namely the stationary point for which $\hat{e}\cdot\hat{n} =\cos\theta_o>0$ (ensuring that the outgoing path actually is outside of the ellipse). Once $\theta^*$ has been found, one can compute the rest of necessary quantities needed in Eq.~\eqref{eq:dpdphi}, namely incidence angle $\cos \theta_i=\hat{\bm{u}}\cdot \hat{\bm{n}}$, refraction angle $\cos \theta_o=\hat{\bm{e}}\cdot \hat{\bm{n}}$, distance from source to boundary $L$ and curvature $\kappa$. An illustrative example of the quantities involved is given in Fig.~\ref{fig:ex_snell}.

It is interesting to check how well the analytical approximation in Eq.~\eqref{eq:v2an}, $v_2^A$, does when compared with the expression obtained as described in this Section, namely $v_2^S$ (which uses Eq.~\eqref{eq:dpdphi}), for the case in which the source is at the center, i.e. $\bm{z}=(0,0)$. Furthermore, we also compare $v_2^S$ for the source at the center versus $v_2^S$ averaged over 60 random source positions. These comparisons are shown in Fig.~\ref{fig:Central&Average_low_e}. The black dots depict $v_2^A$, the solid red line is $v_2^S$ with central source, and the dashed blue line is $v_2^S$ averaged over sources. The eight panels correspond to the eight different centralities explored in Section~\ref{sec:res}, and the potential strength has been set to $V=1$ GeV. The ``nucleus'' size is $R=1$ fm, but as explained in the main text, the results using the SPA are completely independent of the system size.

We observe that the analytical approximation $v_2^A$ already does a pretty good job at describing the actual solution $v_2^S$ for a central source for the more central classes, while sizeable deviations start appearing towards the more peripheral classes. We also see that $v_2^S$ for a central source agrees quite well with the source averaged result, specially at higher values of $k_o$. For this reason, we have chosen to simply show the $v_2^S$ for a central source results as the dashed lines in Fig.~\ref{fig:pot1vsan} in Section~\ref{sec:res}.

\begin{figure}
    \centering
    \includegraphics[width=1\linewidth]{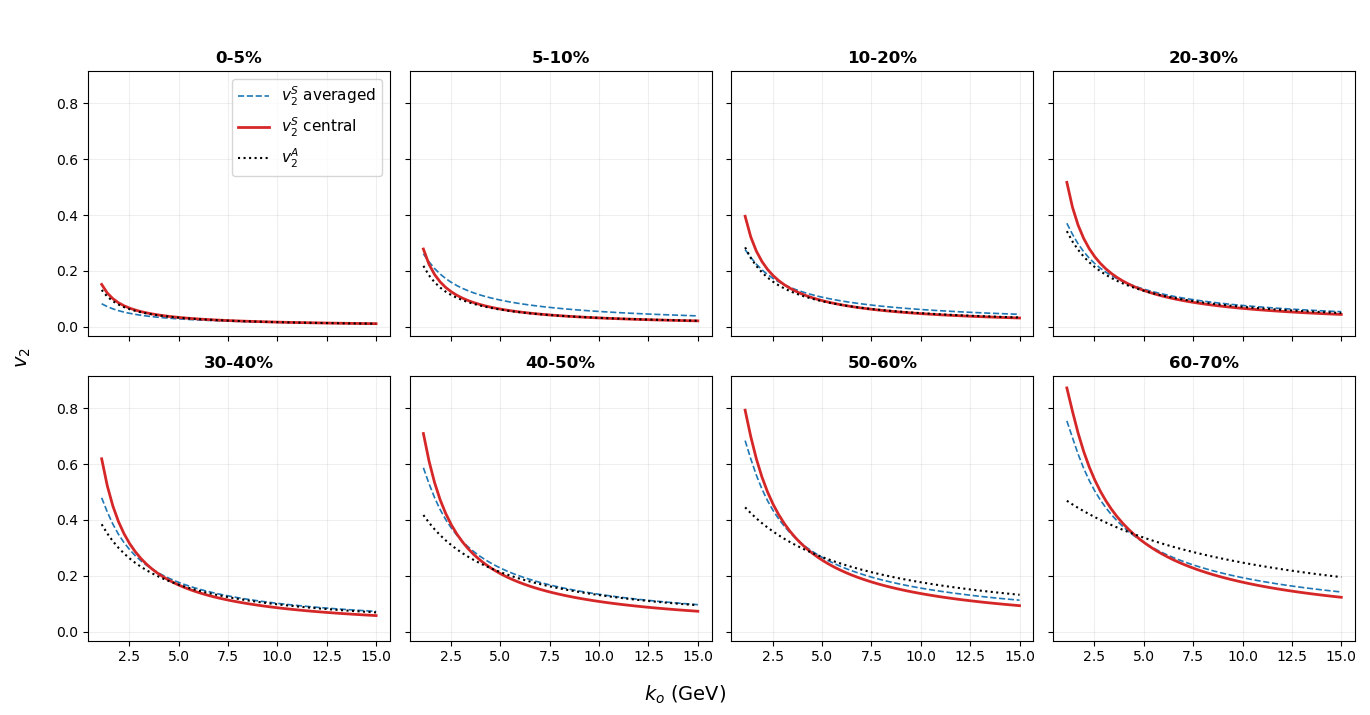}
    \caption{Results for $v_2$ using the SPA as a function of centrality (as defined in Section~\ref{sec:res}) with $V=1$ GeV. We compare $v_2^S$ for a central source (red), $v_2^S$ averaged over 60 random sources (dashed blue) and the analytic approximation for a central source $v_2^A$ (black points).}
    \label{fig:Central&Average_low_e}
\end{figure}

 \clearpage
 \end{widetext}

\end{document}